\documentclass[epj]{svjour}
\usepackage{graphicx,amsmath,ifpdf}

\begin{document}

\hyphenation{e-vents}

\title{Rapidity correlations of protons from a fragmented fireball}
\author{Martin Schulc\inst{1}
   \and Boris Tom\'a\v{s}ik\inst{1,2}}
\institute{
FNSPE, Czech Technical University in Prague, 11519 Prague, Czech Republic
\and
Univerzita Mateja Bela, 97401 Bansk\'a Bystrica, Slovakia
}
\date{\today}

\abstract{We investigate proton rapidity correlations for a fireball that fragments due 
to non-equilibrium effects at the phase transition from deconfined to hadronic phase. 
Such effects include spinodal fragmentation in case of first order phase transition at lower
collision energies, and cavitation due to sudden rise of the bulk viscosity at the crossover
probed at RHIC and the LHC. 
Our study is performed on samples of Monte Carlo events. Correlation 
function in relative rapidity appears to be a sensitive probe of fragmentation. We show that 
resonance decays make the strength of the correlation even stronger.}


\maketitle
%%%%%%%%%%%%%%

\section{Introduction}
\label{intro}

Exploration of properties of very hot strongly interacting matter is the main goal
for which experiments with colliding heavy atomic nuclei at ultrarelativistic 
energies are performed. The suggested phase diagram shows rapid but smooth crossover
from confined to deconfined matter at temperatures around 170~MeV and vanishing baryochemical 
potential \cite{Aoki:2006we}. The smooth crossover becomes a first order phase transition at 
some, presently not precisely known, value of the baryochemical potential \cite{Stephanov:2004wx,Fodor:2004nz}.  
To find the layout of the phase diagram is one of the main goals of  present 
theoretical as well  as experimental activity in the field. 

The fireball which is created in nuclear collisions expands rapidly and so the evolution 
can proceed out of the equilibrium. Particularly,  passage of the bulk through the phase 
transition or the crossover can be too fast and  non-equilibrium scenario could be 
relevant. In case of the first order phase transition this would lead to supercooling \cite{Csernai:1995zn}.
Even maximum possible supercooling and subsequent spinodal decomposition could 
happen \cite{Mishustin:1998eq,Mishustin:2001re,Scavenius:2000bb}. 
There, the bulk becomes mechanically unstable and fragments into droplets of 
high temperature phase. At RHIC and the LHC we likely have the smooth crossover, however. Nevertheless,
in this case calculations show a sharp rise of the bulk viscosity \cite{prattbulk,kharbulk,Meyer:2007dy}
which might lead to 
fireball fragmentation \cite{Torrieri:2007fb,Torrieri:2008ip,Rajagopal:2009yw}, 
as well.  Thus the passage through the phase transition 
can in any case lead to fragmentation of the fireball into smaller droplets which then 
evaporate hadrons.  

Then, the relevant question is: How does one recognise the fragmentation of the bulk?
Correlations of hadrons are a good candidate for an observable which would be sensitive 
to fragmentation. Hadrons originating from the same fragment will have correlated velocities.
Their velocities would be centered around the velocity of the fragment. 
It has been proposed long time ago \cite{pratt94} and reexamined more recently \cite{randrup05} that 
proton correlations are appropriate observables to look at. Protons are rather abundant in 
nuclear collisions and their velocities are close to the velocities of the fragments. This 
is due to the large proton mass which leads to small thermal velocities of protons. Pions,
in comparison, are much more abundant but also very light. Hence, their velocities may differ
considerably from the velocities of the emitting fragments and their correlation function 
is less suitable to study the effect of fragmentation. 

In this paper we study proton correlations in rapidity on datasets simulated with a realistic 
Monte Carlo event generator DRAGON \cite{dragon}. This is an improvement over previous 
simulations \cite{randrup05} where only a schematic distribution of droplets has been used. In our 
model, droplets have sizes distributed according to gamma-distribution and their positions
are random with the only constraint that they must not overlap. It is also important
that resonance decays are included. We
investigate the effect of resonances on proton correlations. 

In the following section we define  the correlation functions which have been measured. In 
Section~\ref{dragon} we give a short overview about DRAGON. Results are presented in 
Section~\ref{res} where we also discuss our predictions for FAIR and the LHC. Conclusions 
are given in Section~\ref{conc}.

%%%%%%%%%%%%%%%%%%%%%%%%%%%%%%%%%%%%%%%%%%%%%%%%%%%%%%%%%%%%%%%%%%%%%%%%%%%%%%%%%%

\section{The correlation function}
\label{cf}

We shall measure the correlation function in 
kinematic variable $\Delta x$ (e.g.\ rapidity difference)
\begin{equation}
C(\Delta x) = \frac{\Phi_2(\Delta x)}{\Phi_{2,\mathrm{mixed}}(\Delta x)}
\end{equation}
where $\Phi_2(\Delta x)$ is the probability to observe a pair of hadrons 
with the distance $\Delta x$. It is averaged over all events of the 
sample.
Denominator $\Phi_{2,\mathrm{mixed}}(\Delta x)$ is the reference 
distribution obtained by the mixed events technique. Hence, 
$\Phi_{2,\mathrm{mixed}}(\Delta x)$ is only determined by the overall distribution 
of hadrons and not affected by any correlations. 

In our study, as $\Delta x$ we shall use the rapidity difference 
\begin{equation}
\label{e:dy1}
\Delta y = y_1 - y_2\, ,
\end{equation}
and the relative rapidity 
\begin{equation}
\label{e:dy3}
y_{12} = \ln \left [ \gamma_{12} + \sqrt{\gamma_{12}^2 - 1} \right ]
\end{equation}
with 
\begin{equation}
\gamma_{12} = \frac{p_1\cdot p_2}{m_1 m_2}\, .
\end{equation}
Note that $y_{12}$ is the rapidity that corresponds to relative velocity of 
one particle in the rest frame of the other
\begin{equation}
y_{12} = \frac{1}{2} \ln \left (\frac{1+v_{12}}{1-v_{12}} \right )\, . 
\end{equation}

In the following we re-derive the correlation function \cite{randrup05}
in $\Delta y$ for the simple case where protons are emitted from droplets 
distributed according to Gaussian distribution. 
The two proton distribution in rapidity in one event is 
written as a sum of two terms, one corresponding to both protons 
coming from the same droplet, other to each proton stemming from 
a different droplet
\begin{equation}
\label{e:r2i}
\rho_2(y_1,y_2) = \sum_i\rho_2^i(y_1,y_2) + \sum_{i\ne j} \rho_1^i(y_1) \rho_1^j(y_2)\, .
\end{equation}
Here, $i,j$ numerate  the droplets and 
$\rho_1^i$ are  the single-particle distributions for production from 
one droplet 
\begin{equation}
\rho_1^i (y) = \frac{\nu_i}{\sqrt{2\pi\sigma^2}} \exp \left ( - \frac{(y - y_i)^2}{2\sigma^2}  \right )\, .
\end{equation}
The droplet is thus centred around $y_i$ and its rapidity distribution has the 
width $\sigma_i$; $\rho_1^i(y)$ is normalized to the total number of protons 
from this droplet $\nu_i$
\begin{equation}
\int \rho_i^i(y)\, dy = \nu_i\, .
\end{equation}
The two-proton distribution for one droplet is then 
\begin{multline}
\rho_2^i(y_1,y_2) = \\
\frac{\nu_i (\nu_i-1)}{{2\pi\sigma^2}}\, 
\exp\left (-\frac{(y_1 - y_i)^2}{2\sigma^2}  \right )\, 
\exp\left (-\frac{(y_2 - y_i)^2}{2\sigma^2}  \right )
\end{multline}
and is normalized to the total number of proton pairs from the 
droplet
\begin{equation}
\int \rho_2^i(y_1,y_2) \, dy_1 \, dy_2 = \nu_i (\nu_i - 1)\, .
\end{equation}
The sums in (\ref{e:r2i}) run over all droplets and over all pairs of droplets in 
the event. 

We next introduce two-particle distribution in \emph{rapidity difference}
in one event
\begin{eqnarray}
\nonumber
F(\Delta y) &=& \int dy_1\, dy_2 \, \delta(\Delta y - (y_1 - y_2))\, \rho_2(y_1,y_2)\\
 & = & 
\int dy_1\, dy_2 \, \delta(\Delta y - (y_1 - y_2))
\nonumber
\\
&& \times \left [ 
\sum_i \rho_2^i (y_1,y_2) + \sum_{i\ne j} \rho_2^i(y_1)\rho_2^j(y_2) 
\right ]\, .
\end{eqnarray}
This leads to
\begin{multline}
\label{e:eqf}
F(\Delta y) = \frac{1}{2} \frac{1}{\sqrt{\pi \sigma^2}} \Biggl [
\exp\left ( - \frac{\Delta y^2}{4\sigma^2} \right ) \sum_i \nu_i (\nu_i -1 ) 
\\
+ 
\sum_{i\ne j} \nu_i \nu_j \exp\left ( - \frac{\Delta y - ( y_i - y_j)}{4\sigma^2} \right )
\Biggr ]\, .
\end{multline}
In order to obtain $\Phi_2(\Delta y)$, this must be now averaged over a large sample 
of events. By doing this we shall also sum over many droplets. In the first 
term on the r.h.s.\ of eq.~(\ref{e:eqf}) this leads to 
\begin{equation}
\frac{1}{N_{\mathrm{ev}}} \sum_{\mathrm{events}\, \alpha} \sum_i \nu_{i,\alpha} ( \nu_{i,\alpha} - 1)
= \langle N_d \rangle \langle \nu (\nu - 1) \rangle_M
\end{equation}
where the subscript $i$ numbers droplets in one event and $\alpha$ stands for different 
events; $\langle N_d \rangle$ is the average number of droplets per event and 
$\langle \dots \rangle_M$ denotes averaging over multiplicity distribution 
for one droplet. Thus $\langle \nu (\nu - 1) \rangle_M$ stands for the average 
number of proton pairs  from one droplet. Assuming further that droplet centres are distributed 
according to Gaussian distribution with the width $\xi$
\begin{equation}
\zeta(y_d) = \frac{1}{\sqrt{2\pi \xi^2}}\, \exp\left ( - \frac{(y_d - y_0)^2}{2\xi^2}\right )
\end{equation}
we can derive that 
\begin{multline}
\label{e:phi2}
\Phi_2(\Delta y) = 
\frac{1}{2} \, \frac{1}{\sqrt{\pi \sigma^2}}\, \langle N_d\rangle \langle \nu(\nu -1)\rangle_M
e^{-\frac{\Delta y^2}{4\sigma^2}}
\\
+ \frac{1}{2}\, \frac{1}{\sqrt{\pi ( \xi^2 + \sigma^2)}} \langle N_d ( N_d - 1) \rangle
\langle \nu \rangle_M^2  e^{-\frac{\Delta y^2}{4 (\xi^2 + \sigma^2)}}\, .
\end{multline}

The denominator of the correlation function is constructed from the mixed pairs sample. 
Hence, there are no pairs where both protons would be emitted from the same droplet. 
The single-particle rapidity distribution is determined by integrating 
over all possible droplet positions
\begin{eqnarray}
\nonumber
\tau(y) & = & \int \frac{1}{\sqrt{2\pi \sigma}}\, \exp\left ( - \frac{(y - \chi)^2}{2\sigma^2} \right )
\\
\nonumber
&& \qquad \qquad \times 
\frac{1}{\sqrt{2\pi \xi^2}} \exp \left ( - \frac{(\chi - y_0)^2}{2\xi^2}\right ) \, d\chi
\\
& = & 
\frac{1}{\sqrt{2\pi ( \xi^2 + \sigma^2)}}\, 
\exp \left ( - \frac{(y - y_0)^2}{2(\xi^2 + \sigma^2)} \right )\, .
\end{eqnarray}
Note that this distribution is normalized to unity. Thus to get 
$\Phi_{2,\mathrm{mixed}}(\Delta y)$ normalized to average number of pairs
per event, we calculate
\begin{multline}
F_{\mathrm{mixed}}(\Delta y)  =  \langle N_d \rangle \langle \nu \rangle_M 
\left ( \langle N_d \rangle \langle \nu \rangle_M - 1 \right ) 
\\
\times \int dy_1 dy_2\, \delta( \Delta y - ( y_1 - y_2))\, \tau(y_1) \tau(y_2)\, .
\end{multline}
Mixed events are by construction averaged over a large number of  events
and thus we can write directly
\begin{multline}
\Phi_{2,\mathrm{mixed}}(\Delta y) = F_{\mathrm{mixed}}(\Delta y)
\\
= \frac{1}{2} \frac{\langle N_d \rangle \langle \nu \rangle_M 
\left ( \langle N_d \rangle \langle \nu \rangle_M - 1 \right )}%
{\sqrt{\pi (\xi^2 + \sigma^2)}}\, \exp\left ( - \frac{\Delta y^2}{4 (\xi^2 + \sigma^2)} \right )\, .
\end{multline}
This is almost identical---up to the term involving multiplicity averages---with the 
second term of the expression for $\Phi_2(\Delta y)$ (\ref{e:phi2}). In practical 
evaluation, however, we shall normalize the correlation function so that it 
converges to unity for large $\Delta y$. This is equivalent to replacing
\begin{equation*}
\langle N_d \rangle \langle \nu \rangle_M 
\left ( \langle N_d \rangle \langle \nu \rangle_M - 1 \right )
\to \langle N_d ( N_d -1)\rangle \langle \nu \rangle_M^2\, .
\end{equation*}
Then we obtain 
\begin{multline}
\label{e:cff}
C(\Delta y) - 1 = \frac{\xi \langle N_d\rangle \langle \nu ( \nu -1 ) \rangle_M}%
{\langle N_d ( N_d -1)\rangle \langle \nu \rangle_M^2}\, 
\sqrt{1 + \frac{\sigma^2}{\xi^2}} \, \frac{1}{\sigma} 
\\
\times 
\exp \left ( - \frac{\Delta y^2}{4\sigma^2 \left ( 1 + \frac{\sigma^2}{\xi^2} \right )} \right )\, .
\end{multline}
Here we see that the correlation function will have the width given by 
$\sigma\sqrt{1+\sigma^2/\xi^2}$. Note that $\xi$ is the width of the droplet
distribution in rapidity, thus the density of droplets will be proportional 
to 
\begin{equation}
n_d \sim \frac{\langle N_d \rangle}{\xi}\, .
\end{equation}
Based on this, for large $\langle N_d\rangle$
\begin{equation}
\frac{\xi\langle N_d \rangle}{\langle N_d (N_d - 1) \rangle} 
\sim \frac{1}{n_d}
\end{equation}
so the magnitude of the correlation function is inversely proportional 
to the rapidity density of the droplets. This is quite natural: 
In a dense configuration of many small droplets---resembling a 
fog---the system becomes homogeneous and correlation function should 
be trivial. 

Note that the choice of Gaussian distribution in this argumentation is 
only due to technical simplicity. Qualitatively, the conclusions 
are more general.

%%%%%%%%%%%%%%%%%%%%%%%%%%%%%%%%%%%%%%%%%%%%%%%%%%%%%%%%%%%%%%%%%%%%%%%%%%%%%%%%%%

\section{The Monte Carlo event generator}
\label{dragon}

Monte Carlo events are prepared with DRAGON: Droplet and hadron generator
for nuclear collisions. Detailed description is available in \cite{dragon}. 
Here we summarize the main features of the model and introduce parameters 
which are varied in subsequent studies. 

Hadrons, including resonances, can be directly produced either from the bulk fireball 
or from the decaying droplets. The fraction of those hadrons emitted from drop\-lets
is a parameter that can be set. Resonances decay according to standard two and 
three-body kinematics with constant matrix element. Chain decays are possible. 
Expected $dN/dy$ or the total multiplicity, which is an experimental observable, 
is among the input parameters. This setting allows for relevant comparisons if 
the multiplicity is kept constant and hadron production is divided between 
bulk and droplet emission in different fractions. 

Directly produced hadrons and the droplets are formed at the usual freeze-out 
hypersurface of the blast wave model. It is characterized by constant longitudinal 
proper time $\tau = \sqrt{t^2 - z^2}$. In this study we only investigate the 
case of central collisions and assume circular transverse profile of the fireball 
with a radius $R$. The fireball also exhibits an expansion pattern which is 
longitudinally boost invariant and has tunable gradient in the transverse direction. 
The expansion velocity profile is given as
\begin{eqnarray}
u^\mu(x) & = & u^\mu (\eta,r,\phi) \nonumber \\
\nonumber
& = &(\cosh \eta\, \cosh \eta_t(r),\, \cos\phi \, \sinh\eta_t(r), \\ 
&&		\qquad					\sin\phi\, \sinh\eta_t(r),\, \sinh\eta \, \cosh\eta_t(r))\\
\eta_t(r) & = & \sqrt{2} \rho_0 \frac{r}{R}\, . 
\end{eqnarray}
where we use the polar coordinates $r$ and $\phi$ in the transverse plane and the space-time rapidity 
\[
\eta = \frac{1}{2}\ln \frac{t+z}{t-z}\, .
\]
These coordinates are taken in the centre of mass system of the collision. 

Droplets are randomly placed within the fireball. Their volumes follow the Gamma distribution
\begin{equation}
{\cal P}_2 (V) = \frac{V}{b^2} e^{-V/b}\, , 
\end{equation}
where the volume parameter $b$ determines the average volume
\[
\overline V = \frac{b}{2}\, .
\]
The masses of the droplets are proportional to their volumes while the energy density 
is taken with the value 0.7~GeV.fm$^{-3}$. Their velocities are given by $u^\mu(x)$ at the 
position where they are produced. Droplets decay exponentially in time with the mean 
lifetime being equal to the radius of the droplet. 

Hadrons, regardless if they come from a droplet or from bulk, are emitted from a thermal 
source with kinetic temperature $T_{\mathrm{kin}}$. In case of production from bulk, the thermal 
distribution is boosted with the local expansion velocity at the point of hadron emission. 
In case of production from  a droplet the thermal distribution is assumed in the rest frame 
of the droplet. 

Chemical composition of the produced hadrons is determined by the assumption of chemical equilibrium 
with temperature $T_{\mathrm{ch}}$. Chemical potentials correspond to conserved quantum numbers
with baryonic chemical potential $\mu_B$ being a free parameter and strangeness chemical potential 
being set  by the requirement of total stran\-ge\-ness neutrality. 

Resonances with masses up to 1.5~GeV/$c^2$ (mesons) and 2~GeV/$c^2$ (baryons) according to the
particle data book are included.

%%%%%%%%%%%%%%%%%%%%%%%%%%%%%%%%%%%%%%%%%%%%%%%%%%%%%%%%%%%%%%%%%%%%%%%%%%%%%%%%%%%

\section{Results}
\label{res}

Correlation functions are evaluated from samples of artificial events. Numerators
are built as histograms including pairs of protons from the same event 
according to their relative rapidity. We investigate rapidity difference
in 1 dimension, eq.(\ref{e:dy1}), and the relative rapidity in three 
dimensions, eq.(\ref{e:dy3}). Denominator of the correlation function is constructed from 
pairs of randomly chosen protons from different events. The constructed 
correlation functions are normalized so that they approach unity for 
large rapidity differences. The obtained data are fitted by Gaussian
form
\begin{equation}
C(x) -1 = A \exp \left ( - \frac{x^2}{2\sigma^2}\right )\, ,
\end{equation}
where $x$ is replaced by either $y_{12}$ or $\Delta y$ and $A$, $\sigma$ 
are fit parameters. 

Most detailed studies are performed on samples of events generated 
with RHIC parameter settings. Chemical composition is characterized 
by $T_{\textrm{ch}} = 170$~MeV and $\mu_B = 46$~MeV. Total hadron rapidity 
density is $dN/dy = 1000$. (Note that this includes also neutral hadrons.)
Kinetic temperature, if not stated otherwise, 
is put to 150~MeV. There is transverse flow with $\rho_0 = 0.7$. 
Rapidity distribution is uniform, but only hadrons from the interval 
$[-1,1]$ are taken into the analysis. For each parameter settings 
a sample of 10,000 events is generated. 

\begin{figure}[t]
\begin{center}
\includegraphics[width=1.0\linewidth]{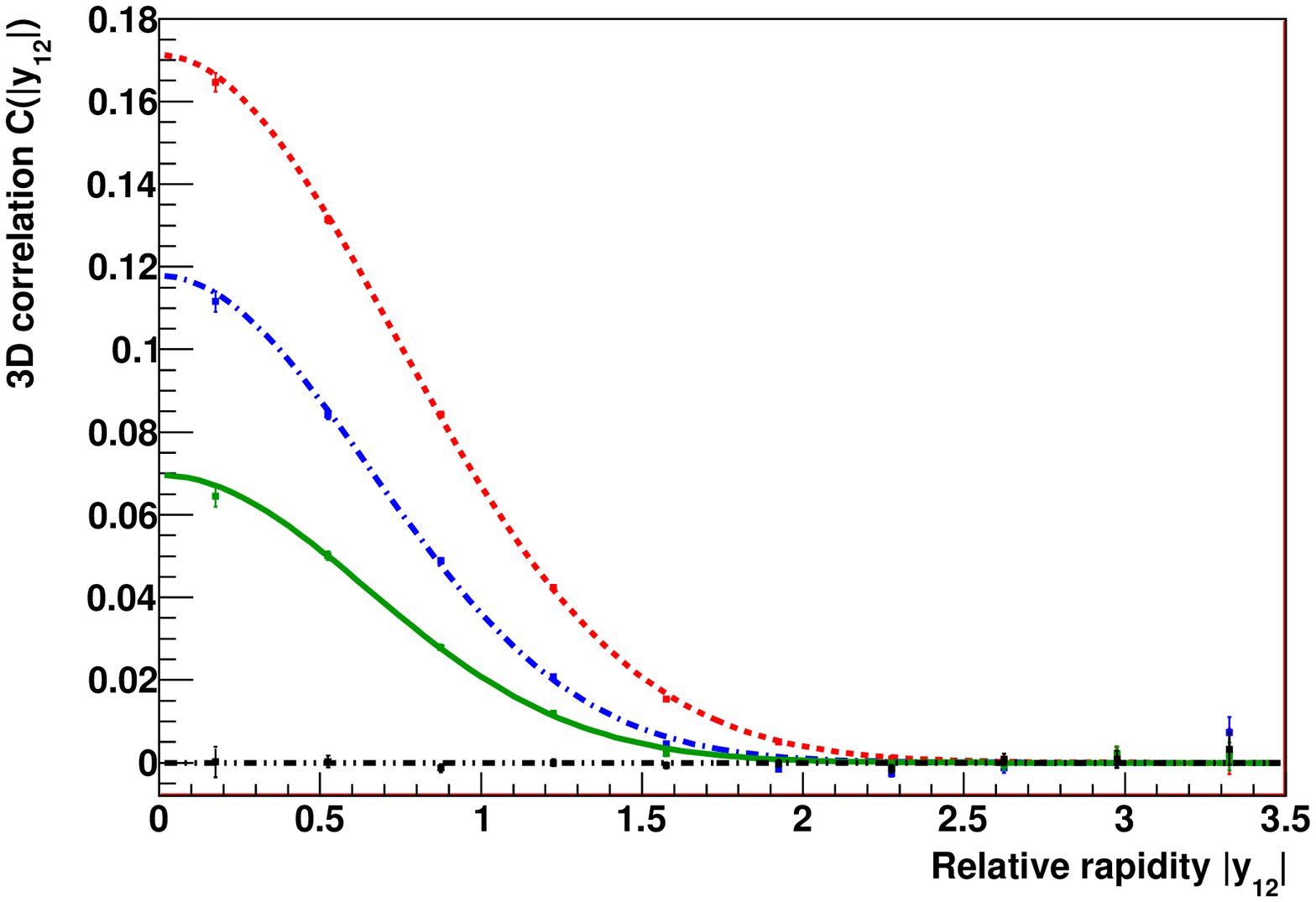}
\includegraphics[width=1.0\linewidth]{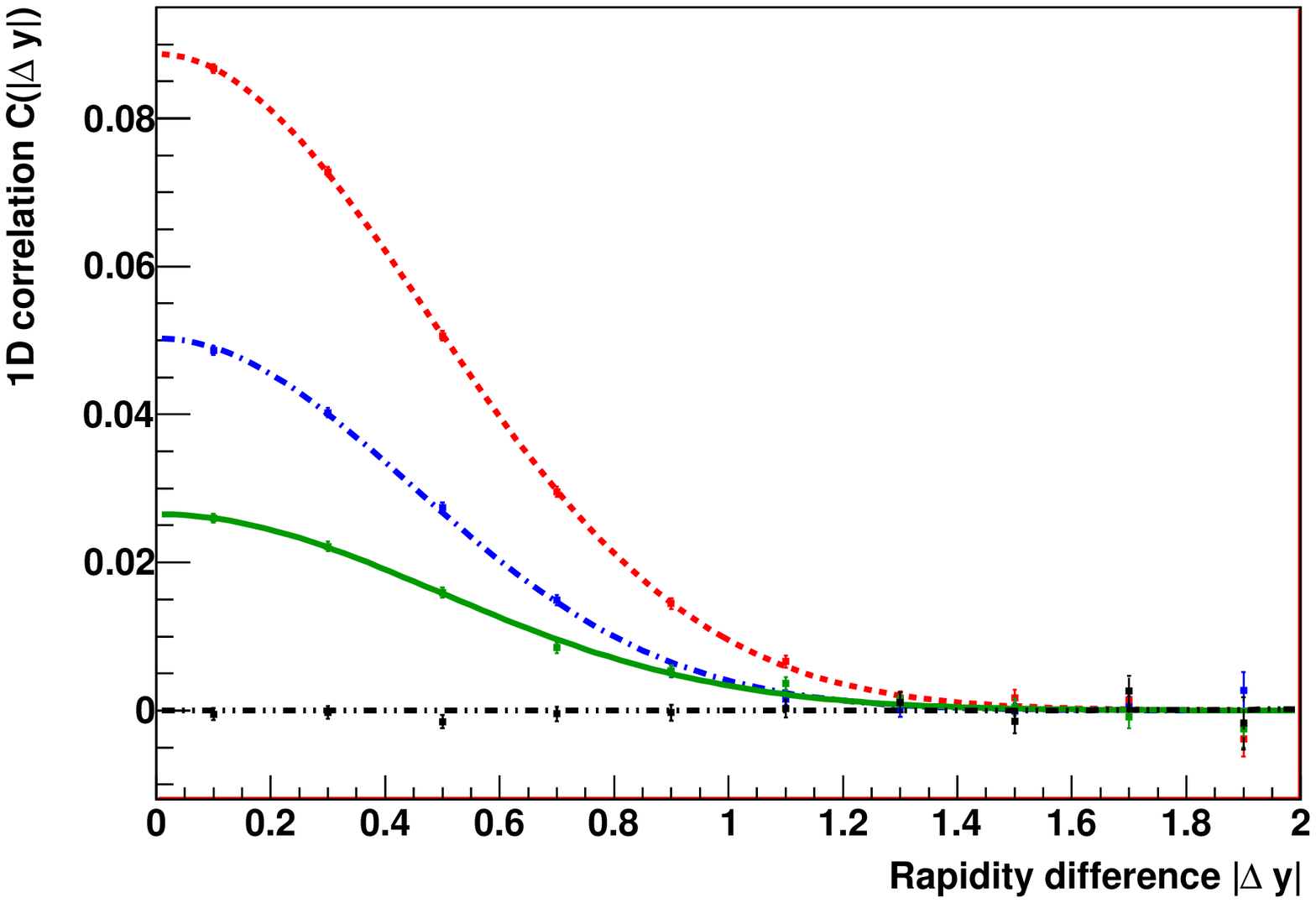}
\end{center}
\caption{
Correlation functions in $y_{12}$ (upper panel) and $\Delta y$
(lower panel)
for RHIC parameter settings with $b=50\,\mathrm{fm}^3$
and varying fraction of droplet-emitted hadrons:
$d = 0$ (black dash-dot-dotted curves), $d = 0.25$
(green solid), $d=0.5$ (blue dash-dotted), and 
$d = 1$ (red dotted).
}
\label{f:d}
\end{figure}

We first consider the case of only \emph{partial} production of hadrons (protons)
from the droplets. In Fig.~\ref{f:d} 
%%%%%%%%%%%%%%%%%%%%%%%%%%%%%%%%%%%%%%%%%%%%%%%%%%%%%%%%%
%%%%%%%%%%%%%%%%%%%%%%%%%%%%%%%%%%%%%%%%%%%%%%%%%%%%%%%%%%
we keep the average size of droplets
constant by fixing $b=50\,\mathrm{fm}^3$ and change the fraction of hadrons 
being emitted by the droplets. 
The other part comes from the bulk fireball while the multiplicity is kept 
constant. Clearly, at vanishing $d$ the correlation function is 
trivial. In both cases, for correlation functions in 1D variable $\Delta y$ 
and 3D variable $y_{12}$, the width is independent of the fraction $d$ 
while the correlation strength $A$ increases as more hadrons come from 
the droplets. We also see that the correlation is about a factor of 2
stronger if plotted as a function of $y_{12}$ (3D correlation).  Therefore,
in the following we shall only study correlation function in $y_{12}$.

\begin{figure}[t]
\begin{center}
\includegraphics[width=1.0\linewidth]{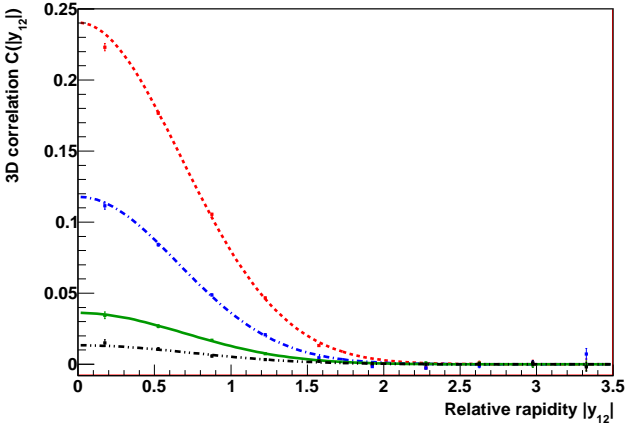}
\end{center}
\caption{
Correlation functions in $y_{12}$
for RHIC parameter settings with one half of hadrons coming from 
droplets, i.e.\ $d = 0.5$. Different curves correspond to different
sizes of the droplets: $b= 10\,\mbox{fm}^3$ (black dash-dot-dotted curve), 
$b= 20\,\mbox{fm}^3$ (green solid), $b= 50\,\mbox{fm}^3$
(blue dash-dotted), and $b= 100\,\mbox{fm}^3$ (red dotted).
}
\label{f:b}
\end{figure}

The effect of varying the size of  droplets is investigated in 
Fig.~\ref{f:b}.
%%%%%%%%%%%%%%%%%%%%%%%%%%%%%%%%%%%%%%%%%%%%%%%%%%%%%%%%%
%%%%%%%%%%%%%%%%%%%%%%%%%%%%%%%%%%%%%%%%%%%%%%%%%%%%%
Here, one half of all hadrons comes from droplets. Since large droplets
correlate more protons, the strength of the correlation function is increased
as $b$ grows. The correlation goes away when the droplets are so small that 
not enough energy is available for a proton pair creation. At $b = 10\, \textrm{fm}^3$ 
the average mass of a droplet is 3.5~GeV which may be enough for two protons to be created,
thus we still see correlation at the level of 0.01. At $b = 5\, \textrm{fm}^3$
this decreases to 1.75~GeV which is insufficient for a proton pair and the 
correlation function would be trivial. 

\begin{figure}[t]
\begin{center}
\includegraphics[width=1.0\linewidth]{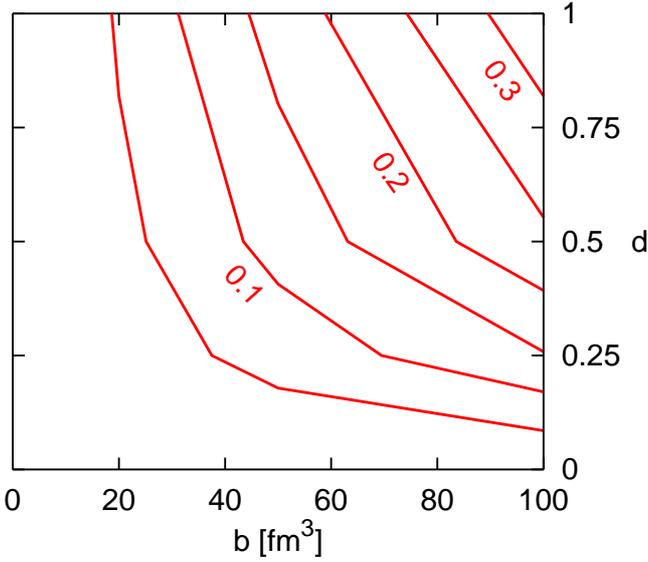}
\end{center}
\caption{
Contour plot shows the dependence of the amplitude $A$ of the 
correlation function calculated for RHIC as a function of droplet
size parameter $b$ and fraction of particles from droplets $d$.
The step between the contours is 0.05, labels show the corresponding 
value of the contours.}
\label{f:A}
\end{figure}

We observe that the correlation strength grows both with bigger droplets and/or
with larger droplet fraction. The dependence of $A$ on  $b$ and $d$ is studied in 
Fig.~\ref{f:A}. 
%%%%%%%%%%%%%%%%%%%%%%%%%%%%%%%%%%%%%%%%%%%%%%%%%%%%%%%%%
%%%%%%%%%%%%%%%%%%%%%%%%%%%%%%%%%%%%%%%%%%%%%%%%%%%%%
Here, we summarize the values which were obtained from fits to 
correlation functions. We conclude that it is not possible to 
determine the droplet size or the droplet fraction just from 
measuring the correlation strength $A$. The value of $A$ can 
at most help to establish a range  of possible combinations 
of $b$ and $d$. If, however, one of them can be determined by some
other means, then rapidity correlations can be used for the 
measurement of the other. 

\begin{figure}[t]
\begin{center}
\includegraphics[width=1.0\linewidth]{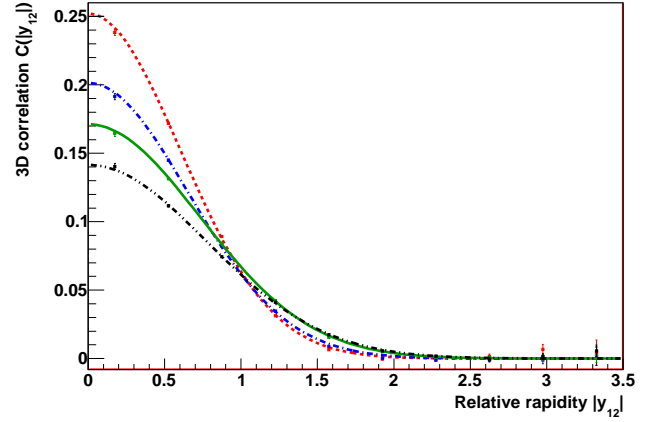}
\end{center}
\caption{
Correlation functions for RHIC parameters with all hadrons 
emitted from droplets and $b = 50~\mbox{fm}^3$. Different 
curves correspond to different kinetic freeze-out temperatures:
$T=170$~MeV  (black dash-dot-dotted curve), $T=150$~MeV 
(green solid), $T=130$~MeV (blue dash-dotted), $T=110$~MeV 
(red dotted).
}
\label{f:T}
\end{figure}

So far we have only investigated the variation of the correlation 
function with changing properties of the drop\-lets.  In 
Fig.~\ref{f:T} 
%%%%%%%%%%%%%%%%%%%%%%%%%%%%%%%%%%%%%%%%%%%%%%%%%%%%%%%%%
%%%%%%%%%%%%%%%%%%%%%%%%%%%%%%%%%%%%%%%%%%%%%%%%%%%%%
we study the effect of temperature variation. 
Average droplet volume is given by $b = 50\,\mathrm{fm}^3$
and all hadrons are emitted from droplets.  We observe 
weakening of the correlation as the temperature increases. 
The interpretation is at hand: the effect of the temperature
is the smearing of hadron velocity so that it does not exactly 
follow that of the droplet. Therefore, lower temperature leads 
to increased correlations. Formally, this effect can be also seen 
in eq.(\ref{e:cff}) (although that equation was derived for $\Delta y$). 
Higher temperature translates into larger $\sigma$.  Through the pre-factor
$1/\sigma$ (at large $\xi$) this makes the correlation weaker.

\begin{figure}[t]
\begin{center}
\includegraphics[width=1.0\linewidth]{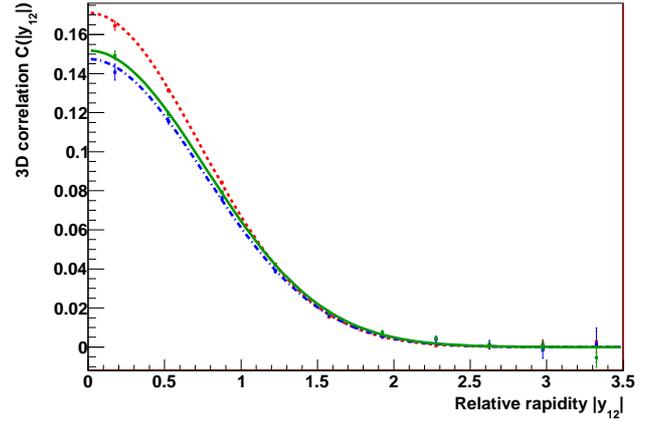}
\end{center}
\caption{
Correlation functions in $y_{12}$ for standard RHIC parameter
settings with all particles emitted from droplets and $b = 50\,\mbox{fm}^3$. 
Different curves show: full correlation function with all resonance decays 
included (red dotted curve), result from a simulation with no resonance production 
included (green solid), result from a simulation with resonances included but 
protons coming from decays of Delta resonances were not taken into analysis
(blue dash-dotted).
}
\label{f:r}
\end{figure}

Our model also allows for dedicated investigation of the effect 
of resonance decays on the observed correlations. This is shown in 
Fig.~\ref{f:r}.
%%%%%%%%%%%%%%%%%%%%%%%%%%%%%%%%%%%%%%%%%%%%%%%%%%%%%%%%%
%%%%%%%%%%%%%%%%%%%%%%%%%%%%%%%%%%%%%%%%%%%%%%%%%%%%%
We see that the resonances increase the intercept of the correlation 
function, i.e., they strengthen the correlation. This is a result 
of two effects: in resonance decay,  daughter particles  
obtain some momentum due to the mass difference between the resonance
and its decay products. This would lead to more smearing of the final 
hadron momenta and lower correlation. On the other hand, resonances are
heavier and therefore closer to the velocity of the emitting droplet 
than lighter hadrons. This increases the correlation. Eventually, the second 
effect wins and the correlation function is increased. Note that the 
smearing of resonance momentum in production from the droplet gets 
weaker if the temperature is lowered. Since in our simulation 
we calculate with rather high kinetic freeze-out temperature of 
150~MeV, it can be expected that this effect will always dominate 
the influence on the correlation function. 

In Fig.~\ref{f:r} we also see that the dominant contribution within 
the resonance decays that produce protons comes from the deltas.

%%%%%%%%%%%%%%%%%%%%%%%%%%%%%%%%%%%%%%%%%%%%%%%%%%%%%%%%%%%%

We next turn to predictions of the proton rapidity correlations 
at other collision energies.
First is that of few GeV per nucleon.  This is the domain 
that should be studied with the help of the planned programs 
at low energy RHIC run, SPS energy scan, FAIR in Darmstadt, and 
NICA at the JINR Dubna. This energy domain seems important since
the critical point of the QCD phase diagram is suspected to 
be here. If so, these accelerators might bring us into the 
region of the phase diagram with first order phase transition 
where spinodal decomposition \cite{Mishustin:1998eq,Mishustin:2001re} and subsequent 
emission of hadrons from droplets can happen. 

For FAIR energy we choose Gaussian profile of the fireball in the space-time 
rapidity $\eta$, with the width $\Delta \eta = 0.7$. Chemical 
composition is taken from the analysis of Pb+Pb collisions 
at CERN SPS at projectile energy 40~$A$ GeV \cite{Becattini:2003wp}: 
chemical freeze-out temperature is 140~MeV 
and the baryochemical potential is 428~MeV. Kinetic freeze-out 
temperature agrees with that of chemical freeze-out. Again, there
is transverse flow with $\rho_0 = 0.4$. The total multiplicity 
is 1250. As a result, on 
average there are 135 protons per event. Results on the 
correlation functions are very similar to those obtained for
RHIC energy and are therefore not shown separately. 

We also make simulations for a fireball as expected in Pb+Pb 
collisions at the LHC. 
Due to near nuclear transparency the baryochemical potential 
is taken at 1~MeV while the chemical freeze-out temperature is 
170~MeV. Kinetic freeze out temperature is set to 150~MeV and
there is slightly higher transverse expansion gradient $\rho_0 = 0.8$. 
Multiplicity density $dN/dy = 2000$.

\begin{figure}[t]
\begin{center}
\includegraphics[width=1.0\linewidth]{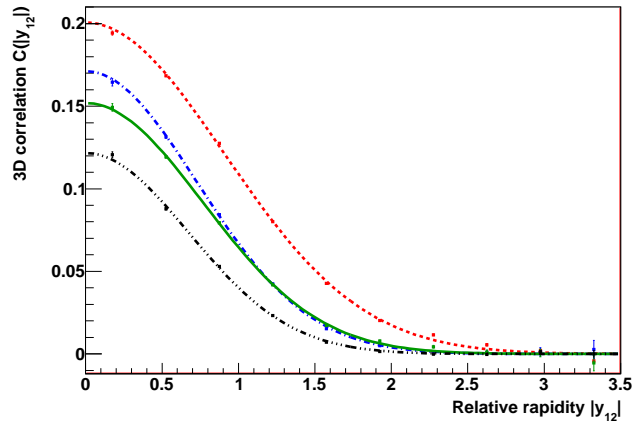}
\end{center}
\caption{
Comparison of correlation functions in $y_{12}$ for different 
collision energies. For parameter sets see text. All hadrons are 
emitted from droplets with $b = 50\,\mbox{fm}^3$. Different curves show: 
FAIR (red dotted curve), RHIC (blue dash-dotted), RHIC without resonances in 
the simulation (green solid), LHC (black dash-dot-dotted). 
}
\label{f:cmp}
\end{figure}

Since results at all investigated collision energies are qualitatively 
very similar, we only show in Fig.~\ref{f:cmp}
%%%%%%%%%%%%%%%%%%%%%%%%%%%%%%%%%%%%%%%%%%%%%%%%%%%%%%%%%
%%%%%%%%%%%%%%%%%%%%%%%%%%%%%%%%%%%%%%%%%%%%%%%%%%%%%
a comparison of the correlation functions at different collision energies. 
The amplitude of the correlation function is highest for FAIR. This is due 
to the regime with lowest \emph{average} $dN/dy$ where clusterization in 
momentum space is best visible. Somewhat surprising, the FAIR correlation 
function is also the widest. We expected just the opposite since the 
kinetic freeze-out energy at FAIR was lower than in the other two cases. 
The larger width of the correlation function is caused by the Gaussian 
profile of the fireball in space time rapidity in comparison with 
uniform profiles at higher energies. Again, we can understand the effect 
to some extent from eq.(\ref{e:cff}): a source with narrow rapidity distribution 
will have smaller $\xi$. This broadens the peak of the Gaussian
correlation function. Paradoxically, fireball that is very narrow in space-time 
rapidity will have wide correlation function.

%%%%%%%%%%%%%%%%%%%%%%%%%%%%%%%%%%%%%%%%%%%%%%%%%%%%%%%%%%%%%%%%%%%%%%%%%%%%%%%

\section{Conclusions}
\label{conc}

Correlation functions in relative rapidity appear to be a powerful tool for the 
study of clustering of hadron production due to fragmentation. Note that in calculating 
the correlation functions we did not include the effect of final state interaction 
(strong and Coulomb) between the protons \cite{pratt94}, which governs 
the correlation function in absence of other effects. This method was also followed in 
\cite{randrup05}. Hence,  our results cannot be directly 
compared to experimental data and should be regarded as an estimate of the size of the 
effect.  It is large indeed, if the droplets have average volume of few tens of fm$^3$. With 
the average size reaching down to the level of a few fm$^3$ the influence on the correlation 
function is on the per cent level. This is the size of the droplets which may be expected 
at RHIC, as indicated by data from PHOBOS.  At the LHC, the droplets, if any, are likely to be 
smaller \cite{Torrieri:2007fb}.

This study assumed that after emission from the droplets, hadrons do not rescatter. 
Rescattering would weaken the correlation and might even wash it away. 

Also, note that we did not study correlation of three and more protons \cite{randrup05}. 
Such correlations demand much better 
statistics and are small if the average volume of droplets drops 
considerably below 10~fm$^3$. For such small droplets 
it becomes unlikely that more than two protons will be emitted from one droplet. Note that 
we do not expect at RHIC droplets much bigger than this (if any). Thus two-particle 
correlation appears to be the most suitable tool at these collision energies.

%%%%%%%%%%%%%%%%%%%%%%%%%%%%%%%%%%%%%%%%%%%%%%%%%%%%%%%%%%%%%%%%%

\paragraph*{Acknowledgements}
We gratefully acknowledge financial support by grants No.~MSM 6840770039, and
LC 07048 (Czech Republic). BT also acknowledges support from
VEGA 1/4012/07 (Slovakia).
%%%%%%%%%%%%%%%%%%%%%%%%%%%%%%%%%%%%%%%%%%%%%%%%%%%%%%%%%%%%%%%%%%

\end{document}